\documentclass[11pt]{article}

\usepackage{amssymb,epsfig,fullpage}

\usepackage{subfigure}
\usepackage{graphicx}

\begin{document}

\title{The effects of symmetry on the dynamics of antigenic variation}
\author{K.B. Blyuss\thanks{Corresponding author. Email: k.blyuss@sussex.ac.uk}\\\\
Department of Mathematics, University of Sussex,\\Brighton, BN1 9QH, United Kingdom}

\maketitle

\begin{abstract}
In the studies of dynamics of pathogens and their interactions with a host immune system, an important role is played by the structure of antigenic variants associated with a pathogen. Using the example of a model of antigenic variation in malaria, we show how many of the observed dynamical regimes can be explained in terms of the symmetry of interactions between different antigenic variants. The results of this analysis are quite generic, and have wider implications for understanding the dynamics of immune escape of other parasites, as well as for the dynamics of multi-strain diseases.
\end{abstract}

\section{Introduction}

In the course of evolution, pathogens have developed various methods of evading the immune system of their hosts. Whilst there are many contributing factors that determine individuals aspects of host-parasite interactions, from a more general perspectives strategies of immune escape can be divided into two major classes. In the first class, parasites remain largely invisible to the immune system of their host through an extended period latency when production of new viruses or bacteria inside the host organism is very small or absent. In this case, the pathogens do not trigger immune response, and thus are able to remain undetected in their hosts for long periods of time. Notable examples of such pathogens include various viruses, such as members of herpes virus family \cite{Jo} and retroviruses \cite{Ba,Ma}. Whilst this method of immune escape is largely unavailable to bacteria due to fundamental differences in replication strategies, several types of bacteria have shown persistence for long periods of time with little evident replication; examples include mycobacteria and {\it T. Pallidum} causing syphilis \cite{L}.

Another possible strategy of immune escape is that, in which pathogen is actively replicating in its host, but this replication is dynamically regulated by the host immune system. There are two ways how this regulation can be achieved: the pathogen can either reach a certain chronic state (equilibrium) through the balance of its proliferation and destruction (see \cite{MT} for {\it Trypanosoma cruzi} example), or it can go through the process of {\it antigenic variation}, whereby it keeps escaping immune response by constantly changing its surface proteins, thus going through a large number of antigenic variants. Perhaps, the best studied pathogens relying on this strategy for immune escape are {\it Plasmodium falciparum} and African Trypanosoma exemplified by {\it Trypanosoma brucei}, with other examples including several families of viruses, bacteria and even fungi \cite{BM,DMW,Gu,HSBS,KHN,Tur}.

In the case of {\it T. brucei}, the organism that causes sleeping sickness, parasite covers itself with a dense homogeneous coat of variant surface glycoprotein (VSG). Genome of {\it T. brucei} has over 1000 genes that control the expression of VSG protein, and switching between them provides the mechanism of antigenic variation \cite{LMRB}. What makes {\it T. brucei}  unique is the fact that unlike other pathogens, whose antigenic variation is typically mediated by DNA rearrangements or transcriptional regulation, activation of VSGs requires recombination of VSG genes into an expression site (ES), which consists of a single {\it vsg} gene flanked by an upstream array of 70 base pair repeats and expression site associated genes (ESAGs). {\it T. brucei} expresses one VSG at any given time, and the active VSG can either be selected by activation of a previously silent ES (and there are up to 20 ES sites), or by recombination of a VSG sequence into the active ES. The precise mechanism of VSG switching has not been completely identified yet, but it has been suggested that the ordered appearance of different VSG variants is controlled by differential activation rates and density-dependent parasite differentiation \cite{LMRB,SSBM}.

For the malaria agent {\it P. falciparum}, the main target of immune response is {\it Plasmodium falciparum} erythrocyte membrane protein-1 (PfEMP1), which is expressed from a diverse family of {\it var} genes, and each parasite genome contains approximately 60 {\it var} genes encoding different PfEMP1 variants \cite{Ga}. The {\it var} genes are expressed sequentially in a mutually exclusive manner, and this switching between expression of different {\it var} gene leads to the presentation of different  variant surface antigens (VSA) on the surface of infected erythrocyte, thus providing a mechanism of antigenic variation \cite{BBMLR,New}. In all cases of antigenic variation, host immune system has to go through a large repertoire of antigenic variants, and this provides parasites with enough time to get transmitted to another host or cause a subsequent infection with a different antigenic variant in the same host. 

Despite individual differences in the molecular implementation of antigenic variation, such as, gene conversion, site-specific DNA inversions, hypermutation etc., there are several features common to the dynamics of antigenic variation in all pathogens. These include ordered and often sequential appearance of parasitemia peaks corresponding to different antigenic variants, as well as certain degree of cross-reactivity. Several mathematical models have been put forward that aim to explain various aspects of antigenic variation. Agur {\it et al.} \cite{Ag} have studied a model of antigenic variation of African trypanosomes which suggests that sequential appearance of different antigenic variants can be explained by fitness differences between single- and double-expressors - antigenic variants that express one or two VSGs. However, this idea is not supported by the experimental evidence arising from normal {\it in vivo} growth and reduced immunogenicity of artificially created double expressors \cite{MJ}. Frank \cite{Fr,FB} has suggested a model that highlights the importance of cross-reactivity between antigenic variants in facilitating optimal switching pattern that provides sequential dominance and extended infection. Antia {\it et al.} \cite{ANA} have considered variant-transcending immunity as a basis for competition between variants, which can promote oscillatory behaviour, but this failed to induce sequential expression. Many other mathematical models of antigenic variation have been proposed and studied in the literature, but the discussion of their individuals merits and limitations is beyond the scope of this work.

In this paper we concentrate a model proposed by Recker {\it et al.} \cite{Re04} (to be referred to as Recker model), which postulates that in addition to a highly variant-specific immune response, the dynamics of each variant is also affected by cross-reactive immune responses against a set of epitopes not unique to this variant. This assumption implies that each antigenic variant experiences two types of immune responses: a long-lasting immune response against epitopes unique to it, and a transient immune response against epitopes that it shares with other variants. The main impact of this model lies in its ability to explain a sequential appearance of antigenic variants purely on the basis of cross-reactive inhibitory immune responses between variants sharing some of their epitopes, without the need to resort to variable switch rates or growth rates (see Gupta \cite{Gu} for a discussion of several clinical studies in Ghana, Kenya and India, which support this theory). 

From mathematical perspective, certain understanding has been achieved of various types of dynamics that can be obtained in the Recker model. In the case when long-lasting immune responses do not decay, numerical simulations in the original paper \cite{Re04} showed that eventually all antigenic variants will be cleared by the immune system, with specific immune responses reaching protective levels preventing each of the variants from showing up again. Blyuss and Gupta \cite{BG} have demonstrated that the sequential appearance of parasitemia peaks during such immune clearance can be explained by the existence of a hypersurface of equilibria in the phase space of the system, with individual trajectories approaching this hypersurface and then being pushed away along stable/unstable manifolds of the saddle-centres lying on the hypersurface. They also numerically analysed robustness of synchronization between individual variants. Under assumption of perfect synchrony, when all variants are identical to each other, Recker and Gupta \cite{RG} have analysed peak dynamics and threshold for chronicity, while Mitchell and Carr \cite{MC} have investigated the additional effect of time delay in the development of immune response. De Leenheer and Pilyugin \cite{LP} have replaced linear growth of antigenic variants in the original model by the logistic growth, and have studied the effects of various types of cross-reactivity on the dynamics, ranging from no cross-reactivity to partial and complete cross-immunty. Mitchell and Carr \cite{MC2} have studied the appearance of synchronous and asynchronous oscillations in the case of global coupling between variants (referred to as "perfect cross immunity" in \cite{LP}).

So far, mathematical analyses of the Recker model have concentrated primarily on identifying and studying different types of behaviour in the model. Whilst this has given a certain headway in the understanding of possible dynamics, symmetric properties of the model have remained largely unstudied, and yet they can provide important insights into the dynamics of the model allowing one to distinguish the interactions between variants and the immune system from the effects of topology of coupling between antigenic variants. The importance of this topology has been highlighted in recent works \cite{BBG,BG10,Re11}. In this paper we study Recker model from the perspective of symmetric dynamical systems. This allows us to perform a systematic analysis of steady states and their stability, as well as to classify various periodic behaviours in terms of their symmetries. The outline of the paper is as follows. In the next section we formalize the Recker model and discuss some of its properties. Section 3 reviews some concepts and techniques from equivariant bifurcation theory. In Section 4 we analyse steady states and their stability with account for symmetry properties of the system. Section 5 is devoted to  symmetry-based classification of different dynamical regimes. The paper concludes in Section 6 with discussion of results.

\section{Mathematical model}

Following Recker {\it et al.} \cite{Re04}, we assume that within a human host, the parasite population of {\it P. falciparum} consists of $N$ distinct antigenic variants, with each antigenic variant $i$, $1\leq i\leq N$, containing a single unique major epitope that elicits a long-lived (specific) immune response, and also several minor epitopes that are not unique to the variant. Assuming that all variants have the same net
growth rate $\phi$, their temporal dynamics is described by the
equation
\begin{equation}\label{yeq}
\frac{dy_i}{dt}=y_i(\phi -\alpha z_i-\alpha' w_i),
\end{equation}
where $\alpha$ and $\alpha'$ denote the rates of variant
destruction by the long-lasting immune response $z_i$ (specific to variant $i$) and by the
transient immune response $w_i$, respectively. The dynamics of the variant-specific immune
response can be written in its simplest form as
\begin{equation}\label{zeq}
\frac{dz_i}{dt}=\beta y_{i}-\mu z_{i},
\end{equation}
with $\beta$ being the proliferation rate and $\mu$ being the
decay rate of the specific immune response. Finally, the transient
(cross-reactive) immune response can be described by the minor
modification of the above equation (\ref{zeq}):
\begin{equation}\label{weq}
\frac{dw_i}{dt}=\beta'\sum_{j\sim i} y_j-\mu' w_i,
\end{equation}
where the sum is taken over all variants $j$ sharing the epitopes with
the variant $i$. We shall use the terms long-lasting and
specific immune response interchangeably, likewise for transient
and cross-reactive.

The above system can be formalized with the help of adjacency or connectivity matrix $A$
whose entries $A_{ij}$ are equal to one if the variants $i$ and $j$ share some of their minor epitopes and
equal to zero otherwise \cite{BG,LP}. Obviously, the matrix $A$ is always a
symmetric matrix. Prior to constructing this matrix it is
important to introduce an ordering of the variants
according to their epitopes. Whilst this choice is pretty arbitrary, it has to be fixed before the analysis can be done.
Consider a system of antigenic variants with just two minor epitopes and two
variants in each epitope. In this case, the total number of variants is four, and we enumerate them as follows
\begin{equation}\label{var4}
\begin{array}{l}
1\hspace{1cm}11\\
2\hspace{1cm}12\\
3\hspace{1cm}22\\
4\hspace{1cm}21
\end{array}
\end{equation}

\begin{figure*}
\hspace{4cm}
\includegraphics[width=8cm]{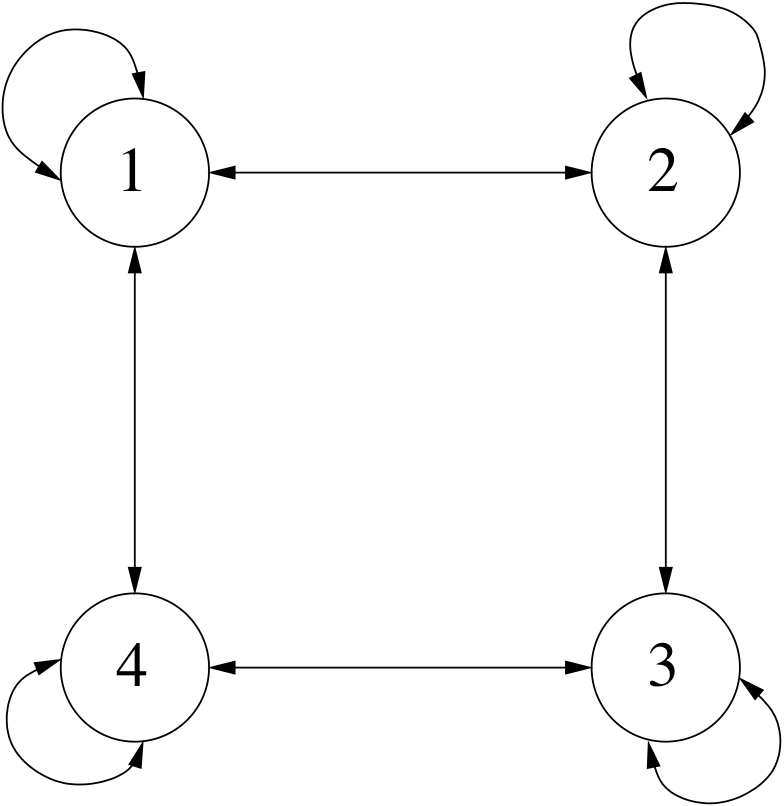}
\caption{Interaction of malaria variants in the case of two minor epitopes with two variants in
each epitope.}\label{vars}
\end{figure*}

After the ordering of variants has been fixed, it is straightforward
to construct the connectivity matrix $A$ of variant
interactions. For the particular system of variants (\ref{var4}) illustrated in Fig.~\ref{vars},
this matrix has the form
\begin{equation}\label{Amat}
A=\left(
\begin{array}{cccccc}
1&1&0&1\\
1&1&1&0\\
0&1&1&1\\
1&0&1&1
\end{array}
\right).
\end{equation}

With the help of lexicographic ordering, one can systematically construct matrix $A$ for an arbitrary number of minor epitopes \cite{BG}.
For the rest of the paper we will concentrate on the
case of two minor epitopes, but the results can be
generalized to larger systems of antigenic variants. Using the
connectivity matrix one can rewrite the system
(\ref{yeq})-(\ref{weq}) in a vector form
\begin{equation}\label{vs}
\frac{d}{dt} \left(\begin{array}{l}{\bf y}\\
{\bf z}\\{\bf w}\end{array}\right) =F({\bf y},{\bf z},{\bf w})=
\left\{
\begin{array}{l}
{\bf y}(\phi{\bf 1}_{N}-\alpha{\bf z}-\alpha'{\bf w}),\\
\beta{\bf y}-\mu{\bf z},\\
\beta'A{\bf y}-\mu'{\bf w},
\end{array}
\right.
\end{equation}
where ${\bf y}=(y_1,y_2,...,y_{N})$, ${\bf z}=(z_1,z_2,...,z_{N})$, ${\bf w}=(w_1,w_2,...,w_{N})$, ${\bf 1}_{N}$ denotes a
vector of the length $N$ with all components equal to one, and in
the right-hand side of the first equation multiplication is taken
to be entry-wise, so that the output is a vector again. The above system
has to be augmented by appropriate initial conditions, which are taken to be
\[
{\bf y}(0)\geq 0,{\bf z}(0)\geq 0,{\bf w}\geq 0.
\]
As it has been shown in \cite{BG}, with these initial conditions the system (\ref{vs}) is well-posed, so its solutions remain non-negative for all time.

We will assume that cross-reactive immune responses develop at a slower rate than specific immune responses, have a shorter life time, and are less efficient in destroying the infection. This implies the following biologically realistic relations between the system parameters
\begin{equation}\label{Pars}
\alpha'\leq \alpha,\hspace{0.5cm}\mu\leq \mu',\hspace{0.5cm}\beta'\leq \beta.
\end{equation}

\section{Elements of equivariant bifurcation theory}\label{EBT}

Before proceeding with the analysis of symmetry effects on the dynamics of systems of antigenic variants, we recall some concepts and results from equivariant bifurcation theory \cite{GSSym,GSS}. Let $\Gamma\subseteq GL(\mathbb{R}^N)$ be a compact Lie group acting on $\mathbb{R}^{N}$. We say that a system of ODEs
\begin{equation}\label{ODE}
\dot{x}=F(x),\hspace{0.5cm}x\in\mathbb{R}^{N},\hspace{0.5cm}F:\mathbb{R}^{N}\to\mathbb{R}^{N},
\end{equation}
is {\it equivariant} with respect to a symmetry group $\Gamma$ if the vector field $F$ commutes with the action of $\Gamma$, i.e. if it satisfies an equivariance condition
\[
F(\gamma x)=\gamma F(x)\hspace{0.3cm}\mbox{ for all }x\in\mathbb{R}^{N},\hspace{0.5cm}\gamma\in\Gamma.
\]
The main examples of the symmetry groups we are interested in are $\Gamma={\bf S}_{n}$, ${\bf Z}_n$ and ${\bf D}_n$ (of order $n!$, $n$, and $2n$, respectively). Here, ${\bf S}_n$ denotes the symmetric group of all permutations in a network with an all-to-all coupling, i.e. all permutations on $n$ symbols. The cyclic group ${\bf Z}_n$ describes the symmetry of a unidirectional ring (rotations only), while the dihedral group ${\bf D}_n$ corresponds to a bidirectional ring of $n$ coupled units (rotations and reflections in the plane that preserve a regular $n$-gon).

One can define the {\it group orbit} of a point $x\in\mathbb{R}^{N}$ under the action of $\Gamma$ is defined as
\[
\Gamma x=\left\{\gamma x\in\mathbb{R}^{N}:\gamma\in\Gamma \right\}.
\]
Note that the equivariance of the system (\ref{ODE}) implies that its equilibria come in group orbits, since if $F(x)=0$ for some $x\in\mathbb{R}^{N}$, then $F(\gamma x)=\gamma 0=0$ for all $\gamma\in\Gamma$. The {\it isotropy subgroup} $\Sigma(x)\subset\Gamma$ of a point $x\in\mathbb{R}^{N}$ is defined as the subgroup of $\Gamma$ which fixes the point $x$, i.e.
\[
\Sigma(x)=\left\{\gamma\in\Gamma:\gamma x=x\right\}.
\]
Associated with each isotropy subgroup $\Sigma\subset\Gamma$ is the {\it fixed-point subspace} denoted by Fix$(\Sigma)$, which is the set of points $x\in\mathbb{R}^{N}$ invariant under the action of $\Sigma$:
\[
{\rm Fix}(\Sigma)=\left\{x\in\mathbb{R}^{N}: \sigma x=x\hspace{0.3cm}\mbox{ for all }\sigma\in\Sigma\right\}.
\]
An important property of fixed-point subspaces is that they are flow-invariant, since if $x\in{\rm Fix}(\Sigma)$, then $\sigma F(x)=F(\sigma x)=F(x)$, and therefore, $F(x)\in{\rm Fix}(\Sigma)$. The equivariant branching lemma states that provided certain (generic) conditions are satisfied by the bifurcation, there exists a branch of equilibrium solutions with symmetry $\Sigma$ for each isotropy subgroup $\Sigma\subset\Gamma$ with ${\rm dim}({\rm Fix}(\Sigma))=1$. Such isotropy subgroups with one-dimensional fixed-point subspaces are called {\it axial}. Similarly, one can define ${\bf C}$-{\it axial} subgroups as those subgroups $\Sigma\subset\Gamma\times {\bf S}^{1}$, for which $\Sigma$ is an isotropy subgroup of the action of $\Gamma\times {\bf S}^{1}$ on the centre subspace of the equilibrium, and  ${\rm dim}({\rm Fix}(\Sigma))=2$.

Next, we recall that a subspace $V$ of $\mathbb{R}^{N}$ is called $\Gamma$-{\it invariant} if
\[
\gamma V\subset V,\hspace{0.5cm}\mbox{ for all }\gamma\in\Gamma.
\]
Two $\Gamma$-invariant subspaces $V$ and $W$ are called $\Gamma$-{\it isomorphic} if there exists a linear isomorphism $T: V\to W$, such that
\[
T(\gamma v)=\gamma (Tv),\hspace{0.5cm}\mbox{ for all }v\in V,\gamma\in\Gamma.
\]
A $\Gamma$-invariant subspace $V$ is $\Gamma$-{\it irreducible} if the only $\Gamma$-invariant subspaces of $V$ are $\{0\}$ and $V$. $\Gamma$-irreducible subspaces can be used to perform an efficient decomposition of the phase space that would allow block-diagonalization of the linearization matrix, thus simplifying the analysis of stability of steady states. The isotypic decomposition proceeds by decomposing $\mathbb{R}^{N}$ into $\Gamma$-irreducible subspaces $V_{j}$ so that $\mathbb{R}^{N}=V_{0}\oplus V_{1}\oplus\cdots\oplus V_{m}$. The {\it isotypic components} are then formed by combining the irreducible subspaces that are $\Gamma$-isomorphic. The {\it isotypic decomposition} is $\mathbb{R}^{N}=W_{0}\oplus W_{1}\oplus\cdots\oplus W_{k}$, $k\leq m$, where the $W_{j}$ are uniquely defined \cite{GSS}.

Now, if $x(t)$ is a $T$-periodic solution of a $\Gamma$-equivariant system (\ref{ODE}), then $\gamma x(t)$ is another $T$-periodic solution of (\ref{ODE}) for any $\gamma\in\Gamma$. Uniqueness of solutions implies that either $x(t)$ and $\gamma x(t)$ are identical, or there exists a phase shift $\theta\in{\bf S}^{1}\equiv\mathbb{R}/\mathbb{Z}\equiv[0,T)$, such that
\[
\gamma x(t)=x(t-\theta).
\]
The pair $(\gamma,\theta)$ is called a {\it spatio-temporal symmetry} of the solution $x(t)$, and the collection of all spatio-temporal symmetries of $x(t)$ forms a subgroup $\Delta\subset\Gamma\times{\bf S}^{1}$. One can identify $\Delta$ with a pair of subgroups, $H$ and $K$, such that $K\subset H\subset\Gamma$. Define
\[
\begin{array}{l}
H\hspace{0.3cm}=\hspace{0.3cm}\left\{\gamma\in\Gamma:\gamma\{x(t)\}=\{x(t)\}\right\}\hspace{0.5cm}\mbox{spatio-temporal symmetries},\\
K\hspace{0.3cm}=\hspace{0.3cm}\left\{\gamma\in\Gamma:\gamma x(t)=x(t)\hspace{0.3cm}\forall t\right\}\hspace{0.6cm}\mbox{spatial symmetries.}
\end{array}
\]
Here, $K$ consists of the symmetries that fix $x(t)$ at each point in time, while $H$ consists
of the symmetries that fix the entire trajectory. Let
\begin{equation}
\displaystyle{L_{K}=\cup_{\gamma\in H\backslash K}{\rm Fix}(\gamma),}
\end{equation}
and let $N_{\Gamma}(K)$ denote the normalizer of $K$ in $\Gamma$:
\[
N_{\Gamma}(K)=\{\gamma\in\Gamma:\gamma K\gamma^{-1}=K\}.
\]
The following theorem gives the necessary and sufficient conditions for $H$ and $K$ to characterize spatio-temporal symmetries of a periodic orbit.\\

\noindent {\bf Theorem} ($H/K$ Theorem \cite{BuG,GSSym}). {\it Let $\Gamma$ be a finite group acting on $\mathbb{R}^{N}$. There is a periodic solution to some $\Gamma$-equivariant systems of ODEs on $\mathbb{R}^{N}$ with spatial symmetries $K$ and spatio-temporal symmetries $H$ if and only if}\\

\noindent (a) $H/K$ {\it is cyclic.}\\
\noindent (b) {\it K is an isotropy subgroup.}\\
\noindent (c) dim Fix$(K)\geq 2$. {\it If }dim Fix$(K)=2$, {\it then either $H=K$ or $H=N_{\Gamma}(K)$.}\\
\noindent (d) {\it H fixes a connected component of} Fix$(K)\backslash L_{K}$.\\

{\it Moreover, when these conditions hold, there exists a smooth $\Gamma$-equivariant vector field
with an asymptotically stable limit cycle with the desired symmetries.}\\\\

The $H/K$ theorem was originally derived in the context of equivariant dynamical systems by Buono and Golubitsky \cite{BuG}, and it has subsequently been used to classify various types of periodic behaviours in systems with symmetry that arise in a number of contexts, from speciation \cite{Spec} to animal gaits \cite{PG} and vestibular system of vertebrates \cite{Vest}.

Now we can proceed with the analysis of symmetry properties of the system (\ref{vs}) as represented by its adjacency matrix $A$. In the case of two minor epitopes with $m$ variants in the first epitope and $n$ variants in the second, the system (\ref{vs}) is equivariant with respect to the following symmetry group \cite{BG}
\begin{equation}
\Gamma=\left\{
\begin{array}{l}
{\bf S}_{m}\times {\bf S}_{n},\mbox{ }m\neq n,\\
{\bf S}_{m}\times {\bf S}_{m}\times{\bf Z}_{2},\mbox{
}m=n.
\end{array}
\right.
\end{equation}
This construction can be generalized in a straightforward way to a larger number of minor epitopes.  It is noteworthy that while within each stratum we have an all-to-all coupling, the full system does not possess this symmetry. System (\ref{vs}) provides an interesting example of a linear coupling, which does not reduce to known symmetric configurations, such as diffusive, star or all-to-all \cite{Pe}. A really important aspect is that two antigenic systems with the same total number of variants $N$ may have different symmetry properties as described by the group $\Gamma$ depending on $m$ and $n$, such that $N=mn$. The simplest example of different kinds of splitting is given by $N=12$, which can be represented as $N=2\times 6$ or as $N=3\times 4$.

\section{Symmetry analysis of steady states}

In the particular case of non-decaying specific immune response $(\mu=0)$, equilibria of the system (\ref{vs}) are not isolated but rather form an $N$-dimensional hypersurface $H_{0}=\{({\bf y},{\bf z},{\bf w})\in\mathbb{R}^{N}:{\bf y}={\bf w=0}_N\}$ in the phase space \cite{BG}. This hypersurface consists of saddles and stable nodes, and in addition to the original symmetry of the system it possesses an additional translational symmetry along the ${\bf z}$ axes. The existence of this hypersurface of equilibria in the phase space leads to a particular behaviour of phase trajectories, which mimics the occurrence of sequential parasitimea peaks in the immune dynamics of malaria \cite{BG,Re04}.

When $\mu>0$, the structure of the phase space of the system (\ref{vs}) and its steady states is drastically different. Now, the only symmetry present is the original symmetry $\Gamma$, and the hypersurface of equilibria $H_{0}$ disintegrates into just two distinct points: the origin $\mathcal{O}$, which is always a saddle, and the fully symmetric equilibrium
\begin{equation}
\begin{array}{l}
E=(Y{\bf 1}_{N},Z{\bf 1}_{N},W{\bf 1}_{N}), \hspace{0.5cm}\mbox{where}\\\\
\displaystyle{
Y=\frac{\phi\mu\mu'}{\alpha\beta\mu'+\alpha'n_{c}\beta'\mu},\hspace{0.5cm}
Z=\frac{\phi\beta\mu'}{\alpha\beta\mu'+\alpha'n_{c}\beta'\mu},\hspace{0.5cm}
W=\frac{\phi\mu n_{c}\beta'}{\alpha\beta\mu'+\alpha'n_{c}\beta'\mu}}.
\end{array}
\end{equation} 
Here $n_{c}$ is the total number of connections for each antigenic variant. It has been previously found by means of numerically computing eigenvalues of the Jacobian that the fully symmetric equilibrium $E$ may undergo Hopf bifurcation as the parameters are varied \cite{BG}. It is worth noting that if one assumes all variants to be exactly the same, the original system (\ref{vs}) collapses to a system with just 3 dimensions, and in this case it is possible to show analytically that the fully symmetric equilibrium is always stable \cite{RG}. This implies that Hopf bifurcation of the fully symmetric equilibrium takes place outside the hyperplane of complete synchrony.

\begin{figure*}
\hspace{2cm}
\includegraphics[width=13cm]{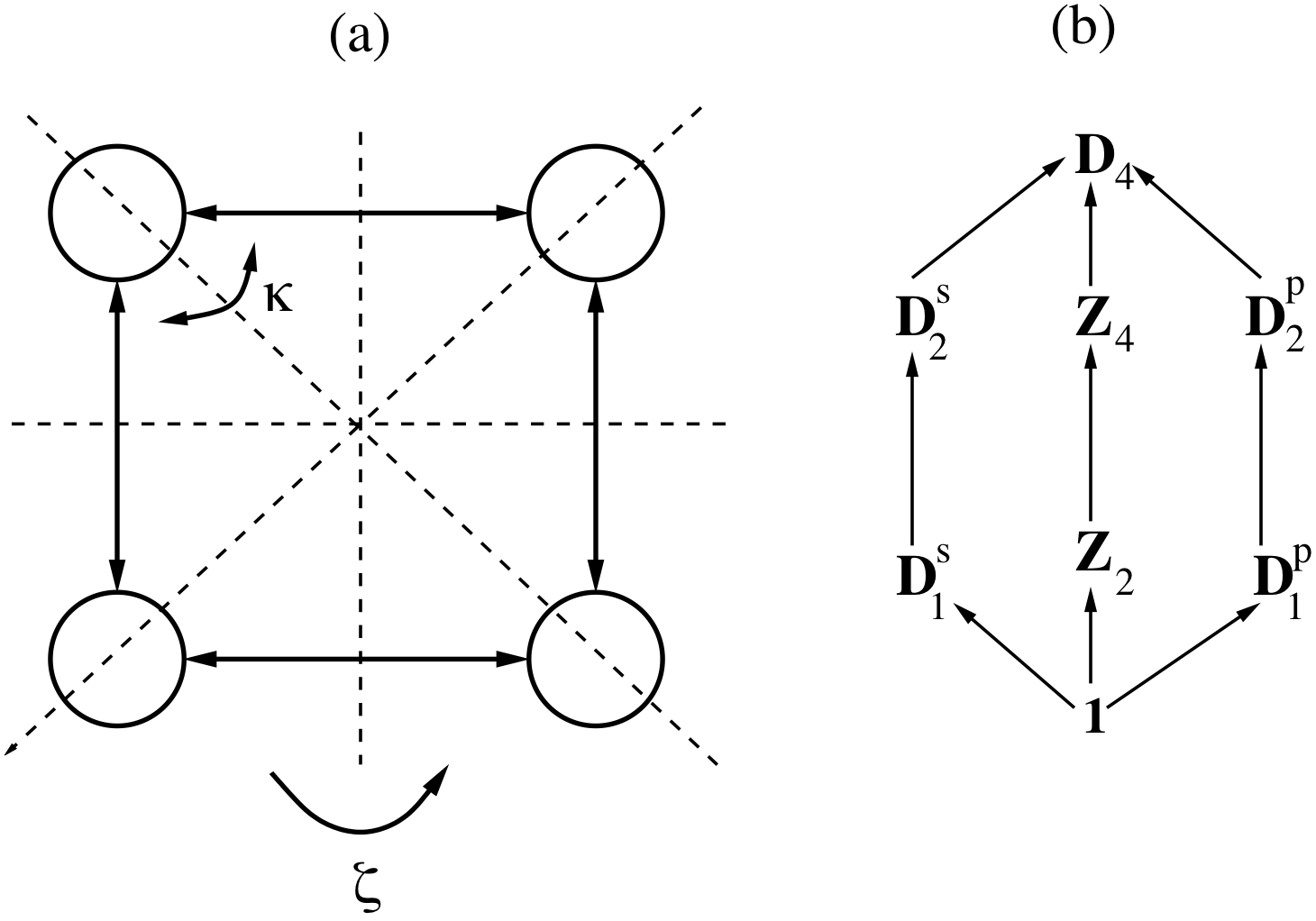}
\caption{(a) Symmetries of the square. (b) Lattice of subgroups of ${\bf D}_4$ symmetry group.}\label{d4sym}
\end{figure*}

In order to find analytically the boundary of Hopf bifurcation in terms of system parameters, as well as to understand the structure of the bifurcating solution, we concentrate on a specific connectivity matrix $A$ that corresponds to a particular case of two epitopes with two variants in each epitope, as described by the system (\ref{var4}). Despite its simplicity, this case is still important for the following two reasons. Firstly, this is the simplest non-trivial combination of antigenic variants which can produce interesting dynamics of interactions with the immune system. Secondly, it can be used as a paradigm for many two-locus two-allele models \cite{GAn,Reck}.

In the case of two epitopes with two variants in each epitope, we have $m=n=2$ and $N=4$, and the system (\ref{vs}) is equivariant under the action of a dihedral group ${\bf D}_{4}$, which is an 8-dimensional symmetry group of a square. This group can be written as ${\bf D}_{4}=\{1,\zeta,\zeta^2,\zeta^3,\kappa,\kappa\zeta,\kappa\zeta^2,\kappa\zeta^3\}$, and it is generated by a four-cycle $\zeta$ corresponding to counterclockwise rotation by $\pi/2$, and a flip $\kappa$, whose line of reflection connects diagonally opposite corners of the square, see Fig.~\ref{d4sym}. The group ${\bf D}_{4}$ has eight different subgroups (up to conjugacy): ${\bf 1}$, ${\bf Z}_{4}$, and ${\bf D}_4$, as well as ${\bf D}_{1}^{p}=\{1,\kappa \}$ generated by a reflection across a diagonal, ${\bf D}_{1}^{s}=\{1,\kappa\zeta\}$ generated by a reflection across a vertical, ${\bf D}_{2}^{p}=\{1,\zeta^2,\kappa,\kappa\zeta^2\}$ generated by reflections across both diagonals, and ${\bf D}_{2}^{s}=\{1,\zeta^2,\kappa\zeta,\kappa\zeta^3\}$ generated by the horizontal and vertical reflections. Finally, the group ${\bf Z}_2$ is generated by rotation by $\pi$. The lattice of these subgroups is shown in Fig.~\ref{d4sym}(b). ${\bf D}_{4}$ has two other subgroups ${\bf Z}_2(\kappa\zeta^2)=\{1,\kappa\zeta^2\}$ and ${\bf Z}_2(\kappa\zeta^3)=\{1,\kappa\zeta^3\}$, which will be omitted as they are conjugate to ${\bf D}_{1}^{p}$ and ${\bf D}_{1}^{s}$, respectively. There is a certain variation in the literature regarding the notation for subgroups of ${\bf D}_4$, and we are using the convention adopted in Golubitsky and Stewart \cite{GSSym}, c.f. \cite{AS,BuG,GSS}.

The group ${\bf D}_4$ has four one-dimensional irreducible representations \cite{FS,GS2}. Equivariant Hopf Theorem \cite{GSSym,GSS} states that under certain genericity hypotheses, there exists a branch of small-amplitude periodic solutions corresponding to each ${\bf C}$-{\it axial} subgroup $\Gamma\times{\bf S}^1$ acting on the centre subspace of the equilibrium. To find out what type of periodic solution the fully symmetric steady state will actually bifurcate to, we can use the subspaces associated with the above-mentioned one-dimensional irreducible representations to perform an isotypic decomposition of the full phase space $\mathbb{R}^{12}$ as follows \cite{DM,Swift}:
\begin{equation}\label{IsD}
\mathbb{R}^{12}=\mathbb{R}^3(1,1,1,1)\oplus \mathbb{R}^3(1,-1,1,-1)\oplus \mathbb{R}^3(1,0,-1,0)\oplus \mathbb{R}^3(0,1,0,-1).
\end{equation}
Jacobian of the linearization of system (\ref{vs}) near any steady state
\[
S=(y_1,y_2,y_3,y_4,z_1,z_2,z_3,z_4,w_1,w_2,w_3,w_4)^{T},
\]
is given by
\begin{equation}\label{Jac}
J(S)=\left(
\begin{array}{cccccccccccc}
P_1&0&0&0&-\alpha y_1&0&0&0&-\alpha' y_1&0&0&0\\
0&P_2&0&0&0&-\alpha y_2&0&0&0&-\alpha' y_2&0&0\\
0&0&P_3&0&0&0&-\alpha y_3&0&0&0&-\alpha' y_3&0\\
0&0&0&P_4&0&0&0&-\alpha y_4&0&0&0&-\alpha' y_4\\
\beta&0&0&0&-\mu&0&0&0&0&0&0&0\\
0&\beta&0&0&0&-\mu&0&0&0&0&0&0\\
0&0&\beta&0&0&0&-\mu&0&0&0&0&0\\
0&0&0&\beta&0&0&0&-\mu&0&0&0&0\\
\beta'&\beta'&0&\beta'&0&0&0&0&-\mu'&0&0&0\\ 
\beta'&\beta'&\beta'&0&0&0&0&0&0&-\mu'&0&0\\ 
0&\beta'&\beta'&\beta'&0&0&0&0&0&0&-\mu'&0\\ 
\beta'&0&\beta'&\beta'&0&0&0&0&0&0&0&-\mu'
\end{array}
\right)
\end{equation}
where
\[
P_i=\phi-\alpha z_i-\alpha' w_i,\hspace{0.5cm}i=1,2,3,4.
\]
For the fully symmetric steady state $E$, this Jacobian takes the block form
\[
J(E)=\left(
\begin{array}{ccc}
{\bf 0}_4 &-\alpha {\bf 1}_4 &-\alpha' {\bf 1}_4\\
\beta {\bf 1}_4 & -\mu {\bf 1}_4& {\bf 0}_4\\
\beta' A & {\bf 0}_4 & -\mu' {\bf 1}_4
\end{array}
\right),
\]
where ${\bf 0}_4$ and ${\bf 1}_4$ are $4\times 4$ zero and unit matrices, and $A$ is the connectivity matrix (\ref{Amat}). Rather than compute stability eigenvalues from this $12\times 12$ matrix, we use the isotypic decomposition (\ref{IsD}) to rewrite this Jacobian in the block-diagonal form \cite{GS2,Swift}
\begin{equation}\label{JacE}
J(E)=\left(
\begin{array}{cccc}
C+2D & {\bf 0}_3&{\bf 0}_3&{\bf 0}_3\\
{\bf 0}_3&C-2D&{\bf 0}_3&{\bf 0}_3\\
{\bf 0}_3&{\bf 0}_3&C&{\bf 0}_3\\
{\bf 0}_3&{\bf 0}_3&{\bf 0}_3&C
\end{array}
\right),
\end{equation}
where
\begin{equation}\label{CD}
C=\left(
\begin{array}{ccc}
0&-\alpha&-\alpha'\\
\beta&-\mu&0\\
\beta'&0&-\mu'
\end{array}
\right),\hspace{0.5cm}
D=\left(
\begin{array}{ccc}
0&0&0\\
0&0&0\\
\beta'&0&0
\end{array}
\right).
\end{equation}
Here, matrix $C$ is associated with self-coupling, and $D$ is associated with nearest-neighbour coupling. Stability changes in the $C+2D$, $C-2D$ and $C$ matrices describe a bifurcation of the fully symmetric steady state $E$ in the even, odd, and $V_4$ subspaces, respectively \cite{Swift}. Prior to performing stability analysis, we recall the Routh-Hurwitz criterion, which states that all roots of the equation
\[
\lambda^3+a_1\lambda^2+a_2\lambda+a_3=0,
\]
are contained in the left complex half-plane (i.e. have negative real part), provided the following conditions hold \cite{Mu}
\begin{equation}\label{RHcon}
\begin{array}{l}
a_i>0,\hspace{0.5cm}i=1,2,3,\\\\
a_1a_2>a_3.
\end{array}
\end{equation}
If the last condition is violated, then the above cubic equation has a pair of purely imaginary complex conjugate eigenvalues when
\begin{equation}\label{RHhopf}
\begin{array}{l}
a_i>0,\hspace{0.5cm}i=1,2,3,\\\\
a_1a_2=a_3,
\end{array}
\end{equation}
as discussed in Farkas and Simon \cite{FaSi}.\\

\noindent {\bf Proposition 1.} {\it The fully symmetric steady state $E$ is stable for} $\alpha'<\alpha'_H$, {\it unstable for }$\alpha'>\alpha'_H$, {\it and 
undergoes a Hopf bifurcation in the odd subspace at} $\alpha'=\alpha'_H$, {\it where}
\begin{equation}
\alpha'_{H}=\frac{\alpha\beta\mu+\mu\mu'(\mu+\mu')}{\beta'\mu'}.
\end{equation}

\vspace{0.2cm}
\noindent{\bf Proof.} Stability of the fully symmetric steady state $E$ changes when one of the eigenvalues of the Jacobian (\ref{JacE}) goes through zero along the real axis or a pair of complex conjugate eigenvalues crosses the imaginary axis. Due to the block-diagonal form of the Jacobian it suffices to consider separately possible bifurcations in the matrices $C$, $C\pm 2D$.

For the matrix $C$ given in (\ref{CD}), the characteristic equation takes the form
\[
\lambda^3+a_1\lambda^2+a_2\lambda+a_3=0,
\]
with
\[
a_1=\mu+\mu',\hspace{0.3cm}a_2=\alpha\beta+\alpha'\beta'+\mu\mu',\hspace{0.3cm}a_3=\alpha\beta\mu'+\alpha'\beta'\mu.
\]
Clearly, in this case $a_{1,2,3}>0$, and also
\[
a_1a_2-a_3=\alpha\beta\mu+\alpha'\beta'\mu'+\mu\mu'(\mu+\mu')>0,
\]
which according to the Routh-Hurwitz conditions (\ref{RHcon}) implies that the eigenvalues of the matrix $C$ are contained in the left complex half-plane for any values of system parameters. This means that the steady state $E$ is stable in the $V_4$ subspace.

Matrix $C+2D$ is equivalent to $C$ upon replacing $\beta'$ with $3\beta'$, which allows one to conclude that the steady state $E$ is also stable in the even subspace for any values of the system parameters.

Finally, for the matrix $C-2D$, the coefficients of the characteristic equation are
\[
a_1=\mu+\mu',\hspace{0.3cm}a_2=\alpha\beta-\alpha'\beta'+\mu\mu',\hspace{0.3cm}a_3=\alpha\beta\mu'-\alpha'\beta'\mu.
\]
Due to biological restrictions on parameters (\ref{Pars}), it follows that $a_{1,2,3}>0$. Computing $a_1a_2-a_3$ gives
\[
a_1a_2-a_3=\alpha\beta\mu+\mu\mu'(\mu+\mu')-\alpha'\beta'\mu'.
\]
When $\alpha'<\alpha'_H$, where
\begin{equation}\label{HopfB}
\alpha'_H=\frac{\alpha\beta\mu+\mu\mu'(\mu+\mu')}{\beta'\mu'},
\end{equation}
we have $a_1a_2-a_3>0$, which according to (\ref{RHcon}) implies that the steady state $E$ is stable in the odd subspace. When $\alpha'=\alpha'_H$, one has $a_1a_2=a_3$, which coincides with the condition (\ref{RHhopf}). Hence, we conclude that at $\alpha'=\alpha'_H$, the steady state $E$ undergoes a Hopf bifurcation in the odd subspace. For $\alpha'>\alpha'_H$, the steady state $E$ is unstable in the odd subspace.\hfill$\blacksquare$\\

The implication of the fact that the Hopf bifurcation can only occur in the odd subspace of the phase space \cite{Swift} is that in the system (\ref{vs}) the fully symmetric state $E$ can only bifurcate to an odd periodic orbit, for which variants 1 and 3 are synchronized and half a period out-of-phase with variants 2 and 4, i.e. each variant is $\pi$ out of phase with its nearest neighbours.

Besides the origin $\mathcal{O}$ and the fully symmetric equilibrium $E$, the system (\ref{vs}) possesses 14 more steady states characterized by a different number of non-zero variants ${\bf y}$ (for a general system with $N$ antigenic variants there would be $2^{N}-2$ of such steady states). To comprehensively analyse these steady states and their stability, let us first introduce auxiliary quantities
\[
Y_{1}=\frac{\phi\mu\mu'}{\beta'\alpha'\mu+\alpha\beta\mu'},
\hspace{0.5cm}Y_{2}=\frac{\phi\mu\mu'}{2\beta'\alpha'\mu+\alpha\beta\mu'},
\]
and
\[
Z_{1,2}=\beta Y_{1,2}/\mu, \mbox{ and }W_{1,2}=\beta' Y_{1,2}/\mu'.
\]
There are four distinct steady states with a single non-zero variant $y_{i}$, which all have the isotropy subgroup ${\bf D}_1^p$ or its conjugate. A representative steady state of this kind is
\begin{equation}\label{E1}
E_1=(Y_1,0,0,0,Z_1,0,0,0,W_1,W_1,0,W_1).
\end{equation}
Other steady states $E_2$, $E_3$ and $E_4$ are related to $E_1$ through elements of a subgroup of rotations ${\bf Z}_4$.\\

\noindent{\bf Proposition 2.} {\it All steady states} $E_1$, $E_2$, $E_3$, $E_4$ {\it with one non-zero variant are unstable.}\\

\noindent{\bf Proof.} As it has already been explained, the steady states $E_{1,2,3,4}$ all lie on the same group orbit. In the light of equivariance of the system, this implies that all these states have the same stability type, and therefore it is sufficient to consider just one of them, for example, $E_1$. Substituting the values of variables in $E_1$ into the Jacobian (\ref{Jac}) gives the characteristic equation for eigenvalues that can be factorized as follows
\begin{eqnarray*}
(\lambda-\phi)&\cdot&(\lambda+\mu)^3(\lambda+\mu')^3(\lambda+\alpha' W_1-\phi)^2\times\\
&&\left[\lambda^3+\lambda^2(\mu+\mu')+\lambda\left(\mu\mu'+Y_1(\alpha\beta+\alpha'\beta')\right)+Y_1(\alpha\beta\mu'+\alpha'\beta'\mu)\right]=0.
\end{eqnarray*}
It follows from this characteristic equation that one of the eigenvalues is $\lambda=\phi>0$ for any values of system parameters, which implies that the steady state $E_1$ is unstable, and the same conclusion holds for $E_{2}$, $E_3$ and $E_4$.\hfill$\blacksquare$\\

Before moving to the case of two non-zero variants, it is worth noting that the symmetry group ${\bf D}_4$ is an example of a more general dihedral group ${\bf D}_{n}$ of order $2n$, for which there are two distinct options in terms of conjugacies of reflections. If $n$ is odd, all reflections are conjugate to each other by a rotation. However, when $n$ is even (as is the case for ${\bf D}_4$), reflections split into two different conjugacy classes: one class contains reflections along axes connecting vertices, and another class contains reflections connecting the sides. These two conjugacy classes are related by an outer automorphism, which can be represented as a rotation through $\pi/n$, which is a half of the minimal rotation in the dihedral group ${\bf D}_{n}$ \cite{GSS}. For the ${\bf D}_4$ group, these two conjugacy classes are reflections along the diagonals, and reflections along horizontal/vertical axes.

Now we consider the case of two non-zero variants, for which there are exactly six different steady states. The steady states with non-zero variants being nearest neighbours on the diagramme (\ref{vars}), i.e. (1,2), (2,3), (3,4) and (1,4), form one cluster:
\[
\begin{array}{l}
E_{12}=(Y_{2},Y_{2},0,0,Z_{2},Z_{2},0,0,2W_{2},2W_{2},W_{2},W_{2}),\\
E_{23}=(0,Y_{2},Y_{2},0,0,Z_{2},Z_{2},0,W_{2},2W_{2},2W_{2},W_{2}),\\
E_{34}=(0,0,Y_{2},Y_{2},0,0,Z_{2},Z_{2},W_{2},W_{2},2W_{2},2W_{2}),\\
E_{14}=(Y_{2},0,0,Y_{2},Z_{2},0,0,Z_{2},2W_{2},W_{2},W_{2},2W_{2}),
\end{array}
\]
while the steady states with non-zero variants lying across each other on the diagonals, i.e. (1,3) and (2,3) are in another cluster
\[
\begin{array}{l}
E_{13}=(Y_{1},0,Y_{1},0,Z_{1},0,Z_{1},0,W_{1},2W_{1},W_{1},2W_{1}),\\
E_{24}=(0,Y_{1},0,Y_{1},0,Z_{1},0,Z_{1},2W_{1},W_{1},2W_{1},W_{1}),
\end{array}
\]
The difference between these two clusters of steady states is in the above-mentioned conjugacy classes of their isotropy subgroups: the isotropy subgroup of the first cluster belongs to a conjugacy class of reflections along the horizontal/vertical axes, with a centralizer given by ${\bf D}_2^s$, and the isotropy subgroup of the second cluster belongs to a conjugacy class of reflections along the diagonals, with a centralizer given by ${\bf D}_2^p$.\\

\noindent{\bf Proposition 3.} {\it All steady states} $E_{12}$, $E_{23}$, $E_{34}$, $E_{14}$, {\it and also } $E_{13}$ {\it and} $E_{24}$, {\it with two non-zero variants are unstable.}\\

\noindent{\bf Proof.} Using the same approach as in {\bf Proposition 2}, due to equivariance of the system and the fact that within each cluster all the steady states lie on the same group orbit, it follows that for the analysis of stability of these steady states it is sufficient to consider one representative from each cluster, for instance, $E_{12}$ and $E_{13}$.

Substituting the values of variables in $E_{12}$ into the Jacobian (\ref{Jac}), one can find the characteristic equation for eigenvalues in the form
\[
\begin{array}{l}
\left(\lambda+\alpha' W_2-\phi\right)^2(\lambda+\mu')^3(\lambda+\mu)^2(\lambda^2+\mu\lambda+\alpha\beta Y_2)\times\\\\
\hspace{0.5cm}\left[\lambda^3+\lambda^2(\mu+\mu')+\lambda\left(\mu\mu'+Y_2(\alpha\beta+2\alpha'\beta')\right)+
Y_2(\alpha\beta\mu'+2\alpha'\beta'\mu)\right]=0.
\end{array}
\]
It follows that this characteristic equation has among its roots an eigenvalue
\[
\lambda=\phi-\alpha'W_2=\frac{\phi(\alpha\beta\mu'+\alpha'\beta'\mu)}{\alpha\beta\mu'+2\alpha'\beta'\mu}>0.
\]
Since this eigenvalue is positive for any values of parameters, we conclude that the steady state $E_{12}$ (and also the steady states $E_{23}$, $E_{34}$, $E_{14}$) is unstable.

In a similar way, the characteristic equation for the steady state $E_{13}$ can be written as follows
\[
\begin{array}{l}
\left(\lambda+2\alpha' W_1-\phi\right)^2(\lambda+\mu)^2(\lambda+\mu')^2\times\\\\
\hspace{0.5cm}\left[\lambda^3+\lambda^2(\mu+\mu')+\lambda\left(\mu\mu'+Y_1(\alpha\beta+\alpha'\beta')\right)+
Y_1(\alpha\beta\mu'+\alpha'\beta'\mu)\right]^2=0.
\end{array}
\]
Hence, one of the eigenvalues is
\[
\lambda=\phi-2\alpha' W_1=\frac{\phi(\alpha\beta\mu'-\alpha'\beta'\mu)}{\alpha\beta\mu'+\alpha'\beta'\mu},
\]
which is always positive for biologically realistic restrictions on parameters (\ref{Pars}). Therefore, we conclude that $E_{13}$ are $E_{24}$ are always unstable.\hfill$\blacksquare$\\

For three non-zero variants, we again have four different steady states, which have an isotropy subgroup ${\bf D}_1^p$ or its conjugate. 
Introducing relative efficacies of the specific and cross-reactive immune responses $E_z$ and $E_w$ \cite{MC,MC2}:
\[
E_z=\frac{\alpha\beta}{\mu},\hspace{0.5cm}E_w=\frac{\alpha'\beta'}{\mu'},
\]
one can write the values of state variables for these states in terms of  $E_z$ and $E_w$ as follows:
\begin{equation}\label{TNZ}
\begin{array}{l}
\displaystyle{y_1=\frac{E_z-E_w}{E_z}Y,\hspace{0.3cm}
y_2=y_4=Y,\hspace{0.3cm}y_3=0,\hspace{0.3cm}Y=\frac{\phi E_z}{E_z^2+2E_zE_w-E_w^2},}
\\\\
\displaystyle{z_1=\frac{\beta}{\mu}\frac{E_z-E_w}{E_zi}Y,\hspace{0.3cm}z_2=z_4=\frac{\beta}{\mu}Y,\hspace{0.3cm}z_3=0,}
\\\\
\displaystyle{w_1=\frac{\beta'}{\mu'}\left(2+\frac{E_z-E_w}{\phi}\right)Y,\hspace{0.5cm}
w_2=w_4=\frac{\beta'}{\mu'}\left(1+\frac{E_z-E_w}{E_z}\right)Y,\hspace{0.5cm}w_3=2\frac{\beta'}{\mu'}Y.}
\end{array}
\end{equation}
Here, we have chosen the three non-zero variants to be variants $1$, $2$ and $4$. The other three steady states can be obtained from this one by rotation of variants.\\

\noindent{\bf Proposition 4.} {\it All steady states with three non-zero variants are unstable.}\\

\noindent{\bf Proof.} Similar to the proofs of {\bf Proposition 2} and {\bf 3}, we employ system equivariance and the fact that the steady states with three non-zero variants all lie on the same group orbit to conclude that they all have the same stability type. Substituting the values of variables in the steady state (\ref{TNZ}) into the Jacobian (\ref{Jac}) yields the characteristic equation
\[
\left(\lambda-\phi+2\frac{\alpha'\beta'}{\mu'}Y\right)(\lambda+\mu)(\lambda+\mu')P_{9}(\lambda)=0,
\]
where $P_{9}(\lambda)$ is a $9$-th degree polynomial in $\lambda$. It follows that one of the characteristic eigenvalues is
\[
\lambda=\phi-2\frac{\alpha'\beta'}{\mu'}Y=\frac{\phi\left(\alpha^2\beta^2\mu'^2-\alpha'^2\beta'^2\mu^2\right)}{\alpha^2\beta^2\mu'^2-\alpha'^2\beta'^2\mu^2+2\alpha\beta\mu\alpha'\beta'\mu'},
\]
which is always positive due to biological restrictions on parameters (\ref{Pars}). Hence, the steady state (\ref{TNZ}) and the other three steady states with three non-zero variants are all unstable.\hfill$\blacksquare$\\

\begin{figure*}
\hspace{2cm}
\includegraphics[width=13cm]{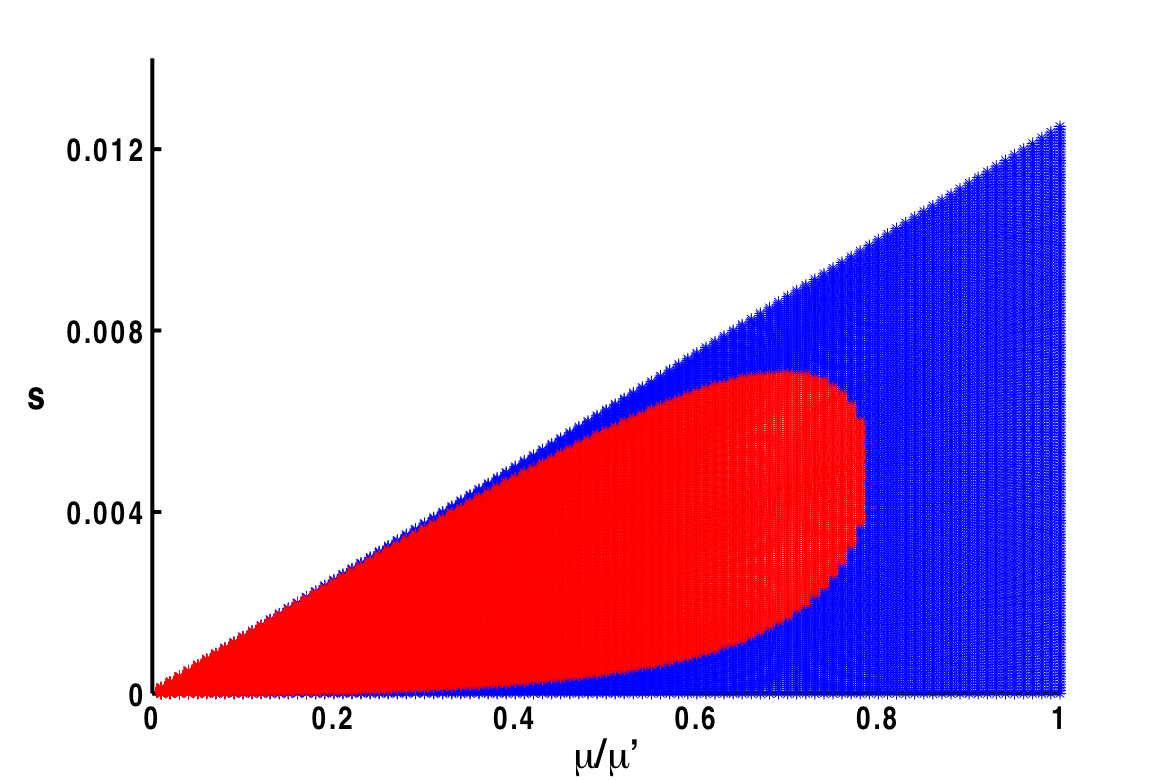}
\caption{Lines of equilibria in the system  (\ref{vs}) for $E_z=E_w$. Each vertical line is parameterized by $s$. Parameter values are $\phi=7.5$, $\alpha=15$, $\beta=\beta'=10$, $\mu'=0.5$}\label{LEQ}
\end{figure*}

This completes the analysis of the steady states of system (\ref{vs}) with ${\bf D}_4$ symmetry for generic values of parameters. It is worth noting, however, that in a particular case when $E_z=E_w$, the determinant of the system of linear equations that describes the steady states is equal to zero, and this results in the existence of a line of equilibria
\[
y_{1}=y_{3}=\frac{\phi\mu}{2\alpha\beta}-s,\hspace{0.5cm}y_{2}=y_{4}=s,\hspace{0.5cm}0<s<\frac{\phi\mu}{2\alpha\beta}.
\]
parameterized by an additional free parameter $s$. This line is reminiscent of the hypersurface of equilibria in the case $\mu=0$ in that every point on each such line is a steady state of the system (\ref{vs}), and as one moves along the line of equilibria, stability of the steady states can change, as illustrated in Fig.~\ref{LEQ}. One can observe that as the decay rate of the specific immune responses increases, this leads to stabilization of such steady states, and provided this rate is sufficiently high, all steady states on lines of equilibria are stable.

\begin{figure*}
\hspace{0.5cm}
\includegraphics[width=15cm]{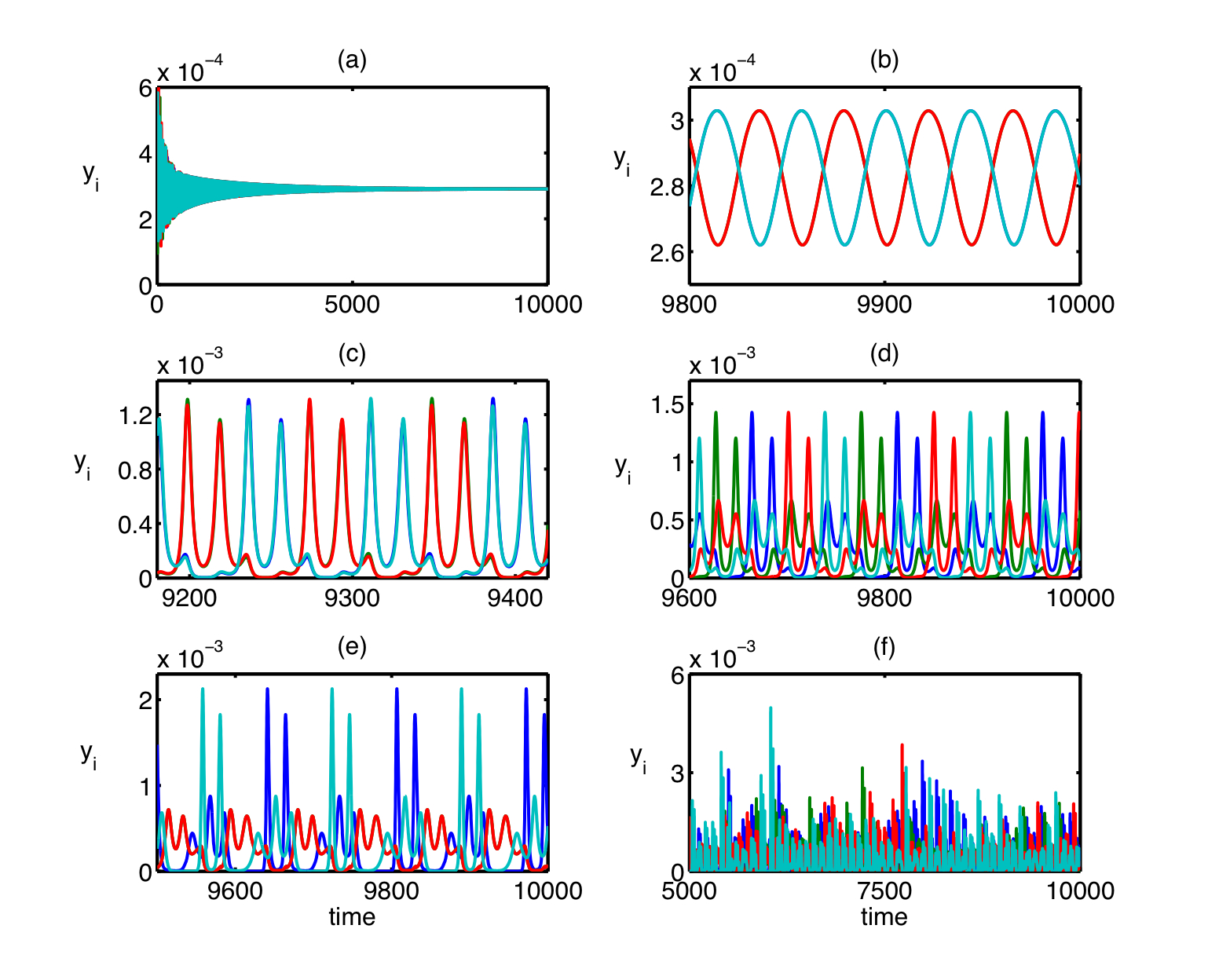}
\caption{Temporal dynamics of the system (\ref{vs}). Parameter values are $\phi=7.5$, $\beta=10$, $\beta'=9$, $\alpha=15$, $\mu=0.01$, $\mu'=0.02$. (a) Stable fully symmetric equilibrium ($\alpha'=8$). (b) Anti-phase periodic solution ($\alpha'=8.5$), spatio-temporal symmetry $(H,K)=({\bf D}_4,{\bf D}_{2}^{p})$. (c) Anti-phase periodic solution ($\alpha'=11.6$), spatio-temporal symmetry $(H,K)=({\bf D}_4,{\bf D}_{2}^{p})$. (d) Discrete rotating wave ($\alpha'=11.6$), spatio-temporal symmetry $(H,K)=({\bf Z}_4,{\bf 1})$. (e) Out-of-phase oscillations ($\alpha'=11.6$), spatio-temporal symmetry $(H,K)=({\bf D}_2^p,{\bf D}_1^p)$. (f) Chaos ($\alpha'=12$).}\label{dyn_fig}
\end{figure*}

\section{Dynamical behaviour of the model}

In the previous section we established that a fully symmetric steady state $E$ can undergo Hopf bifurcation, giving rise to a stable anti-phase periodic solution. Now we look at the evolution of this solution and its symmetries under changes in system parameters. For convenience, we fix all the parameters except for the rate of variant destruction by transient immune response $\alpha'$, which is taken to be a control parameter. The results of numerical simulations are presented in Fig.~\ref{dyn_fig}. When $\alpha'$ is sufficiently small to satisfy the condition $\alpha'<\alpha'_H$, the fully symmetric steady state is stable, as shown in plot (a). As it crosses the boundary $E_{w}=E_{z}$, the fully symmetric steady state loses stability through a Hopf bifurcation giving rise to a periodic solution with the spatio-temporal symmetry $H={\bf D}_4$ and a spatial symmetry $K={\bf D}_{2}^{p}$, which ensures that the variants 1 and 3 have the same behaviour, as do variants 2 and 4, as is illustrated in figure (b). Figure (c) indicates that as $\alpha'$ increases, the periodic solution retains its symmetry but changes temporal profile, acquiring two peaks during a single period.

When analysing other types of periodic behaviour in the model, one can note that according to the $H/K$ theorem (see Section~\ref{EBT}), periodic states can have spatio-temporal symmetry group pairs $(H,K)$ only if $H/K$ is cyclic, and $K$ is an isotropy subgroup \cite{BG,GSSym}. In the case of ${\bf D}_4$ symmetry group acting on four elements, there are eleven pairs of subgroup $H$ and $K$ satisfying these requirements \cite{GSSym}. Now we look at actual solutions of the model and identify these spatial and spatio-temporal symmetries. For the same value of $\alpha'$ as in Fig. (c), the system exhibits two more solutions with quite different spatial and spatio-temporal symmetries, as demonstrated in Figs. (d) and (e). The solution shown in Fig. (d) has the symmetry $(H,K)=({\bf Z}_4,{\bf 1})$ and is a discrete travelling wave, also known as a "splay state" \cite{SM}, "periodic travelling (or rotating) wave" \cite{AKS}, or "ponies on a merry-go-round" or POMs \cite{AGM} in the studies of systems of coupled oscillators. In this dynamical regime all variants appear sequentially one after another along the diagramme (\ref{vars}) with quarter of a period difference between two neighbouring variants. From the perspective of equvariant bifurcation theory, this solution is generic since the group ${\bf Z}_{n}$ is always one of the subgroups of the ${\bf D}_n$ group for the ring coupling,  or the ${\bf S}_n$ group for an all-to-all coupling, and its existence has already been extensively studied \cite{AGM,GS2,GSS}. From the immunological point of view, this is an extremely important observation that effectively such solution, which immunologically represents sequential appearance of parasitemia peaks corresponding to different antigenic variants, owes its existence not to the individual dynamics of antigenic variants, but rather to the particular symmetric nature of cross-reactive interactions between them. This immunological genericity ensures that the same conclusions hold for a wide variety of immune interactions between human host and parasites, which use antigenic variation as a mechanism of immune escape, as illustrated, for instance, by malaria, African Trypanosomes, several members of {\it Neisseria} family ({\it N. meningitidis} and {\it N. gonorrhoeae}), {\it Borrelia hermsii} etc. \cite{Gu,Tur}.

Another co-existing solution for the same value of $\alpha'$ is a state shown in Fig. (e). This solutions is characterized by the symmetry $(H,K)=({\bf D}_2^p,{\bf D}_1^p)$ and corresponds to a situation, in which variants 1 and 3 are oscillating half a period out-of-phase with each other, and the variants 2 and 4 coincide and oscillate at twice the frequency of the pair (1,3). As the value of $\alpha'$ is increased further, the system demonstrates a state of deterministic chaos, when variants appear in arbitrary order and magnitude. These kinds of periodic and chaotic solutions have been previously identified as most relevant for the clinical analysis of blood-stage malarial infection \cite{RM}.

\section{Discussion}

In this paper we have used techniques of equivariant bifurcation theory to perform a comprehensive analysis of steady states and periodic solutions in a model of antigenic variation in malaria. In the simplest case of two epitopes with two variants in each epitope, the system is equivariant with respect to a ${\bf D}_4$ symmetry group of the square. Using isotypic decomposition of the phase space based on the irreducible representations of this symmetry group has allowed us to find a closed form expression for the boundary of Hopf bifurcation of the fully symmetric steady state in terms of system parameters, as well as to show that for any parameter values this bifurcation leads to an anti-phase periodic solution. All other steady states have been classified in terms of their isotropy subgroups, and this has provided insights into their origin and relatedness. With the help of the $H/K$ theorem, we have classified various periodic states of the model in terms of their spatial and temporal symmetries.

One of the important issues that have to be taken into account when applying methods of equivariant bifurcation theory to the models of realistic systems is the fact that these systems do not always fully preserve the assumed symmetry. In the context of modelling immune interactions between distinct antigenic variants, this means that not all variants cross-react in exactly the same quantitative manner. Despite this limitation, due to the normal hyperbolicity, which is a generic property in such models, main phenomena associated with the symmetric model survive under perturbations, including symmetry-breaking perturbations. The discussion of this issue in the context of modelling sympatric speciation using a symmetric model can be found in \cite{GSSym}.

Whilst the analysis in this paper was constrained to the models of within-host dynamics antigenic variation, a similar approach can also be used for the population level mathematical models of multi-strain diseases. These models usually have a similar structure in that all strains/serotypes are assumed to be identical and have a certain degree of cross-protection or cross-enhancement based on antigenic distance between them \cite{AF,AdSa,CKM,DG,GG,GMN,GFA,MF,RG05,Reck}. The fact that multiple strains are antigenically related introduces symmetry into a model, and this then transpires in different types of periodic behaviours observed in these models. To give one example, almost all of these models support a solution in the form of discrete travelling wave having a symmetry $(H,K)=({\bf Z}_n,{\bf 1})$, which represents sequential appearance of different strains \cite{GFA,MF,Reck}. In a model studied by Calvez {\it et al.} \cite{CKM} (based on an earlier model of Gupta {\it et al.} \cite{GFA}), the authors study a multi-strain model having symmetry of a cube, and they find that the fully symmetric steady state can undergo Hopf bifurcation and give rise to an antiphase solution with a tetrahedral symmetry, which was not expected {\it a priori}. At the same time, if one looks at this system from the symmetry perspective, such periodic solution is to be expected as it has a ${\bf D}_4$ symmetry, which corresponds to one of the three maximal isotropy subgroups and is therefore generically expected to arise at a Hopf bifurcation \cite{Fie,JLV}. Observations of this kind illustrate that taking into account symmetries of the underlying models of multi-strain diseases can help make significant inroads in understanding and classifying possible periodic behaviours in such systems.

The results presented in this paper are quite generic, and the conclusions we obtained are valid for a wide range of mathematical models of antigenic variation. In fact, they are applicable to the analysis of within-host dynamics of any parasite, which exhibits similar qualitative features of immune interactions based on the degree of relatedness between its antigenic variants. The significance of this lies in the possibility to classify expected dynamical regimes of behaviour using very generic assumption regarding immune interactions, and they will still hold true provided the actual system preserves the underlying symmetries. In the model studied in this paper the degree of cross-reactivity between antigenic variants does not vary with the number of epitopes they share. It is straightforward, however, to introduce antigenic distance between antigenic variants in a manner similar to the Hamming distance \cite{AdSa,RG05,SFAP}. Such a modification would not alter the topology of the network of immune interactions but rather assign different weights to connections between different antigenic variants in such a network. Symmetry analysis of the effects of antigenic distance on possible dynamics are the subject of further study.

\section*{Acknowledgments} The author would like to thank Peter Ashwin, Jon Dawes and Martin Golubitsky for useful discussions. The author would also like to thank an anonymous referee for helpful comments and suggestions.

\end{document}